\documentclass[aps,pre,twocolumn,showpacs]{revtex4b5}
\usepackage{amssymb,graphicx,float,isolatin1}
\newcommand{\cbeta}{{\cal B}}

\begin{document}
\title{The Origins of Phase Transitions in Small Systems}
\author{Oliver M{\"u}lken, Heinrich Stamerjohanns, and Peter Borrmann}
\affiliation{Department of Physics, Carl von Ossietzky University
Oldenburg, D-26111 Oldenburg, Germany}
\date{\today}
\begin{abstract}
The identification and classification of phases in small systems,
e.g. nuclei, social and financial networks, clusters, and biological
systems, where the traditional definitions of phase transitions are 
not applicable, is important to obtain a deeper understanding of the
phenomena observed in such systems. 
Within a simple statistical model we investigate the validity and
applicability of different classification schemes for phase transtions in
small systems. We show that the whole complex temperature plane contains
necessary information in order to give a distinct classification.
\end{abstract}
\maketitle

The thermodynamics of small systems, e.g. Bose-Einstein condensates 
in magneto-optical traps\cite{Anderson:1995,Bradley:1995,Davis:1995},  
the nuclear liquid-gas transition observed by multifragmentation in heavy ion 
reactions ~\cite{Nebauer:1999,Pochodzalla:1995,Ma:1997}, and 
the solid-liquid phase transition of sodium
clusters~\cite{Schmidt97,Schmidt98},
have gained increasing interest over the last few years.
Because these systems are far away from the thermodynamic 
limit the standard tools for the description of phase transitions 
are not applicable and new concepts are needed. 
Within the last few years several classification schemes for phase
transitions in finite systems have been proposed~\cite{Dgross:2000}.
In this Letter we compare these classification schemes by means of 
a simple statistical models for atomic clusters and show that graveling 
transitions occurring in these models can only completely understood
by considering the whole complex temperature plane.

Among others Gross {\sl et al.}~have suggested a microcanonical
treatment~\cite{Gross:1997,Dgross:1997,Dgross:2000}, where phase
transitions of different order are distinguished by the curvature of the
entropy $S= k_{\rm B} \ln\Omega(E)$.  According to their scheme a {\sl
back-bending} in the microcanonical caloric curve
$T(E)=1/\partial_E\ln(\Omega(E))$, i.e. the appearance of  negative heat
capacities, is a mandatory criterion for a first order transition.
Caloric curves without back-bending, where the associated specific heat
shows a hump, are classified as higher order transitions.

We have proposed a classification scheme based on the distribution of
zeros of the canonical partition function in the complex
temperature plane~\cite{Borr99b}. The classical partition function 
\begin{equation} 
Z(\beta) = \left( \frac{1}{2 \pi \beta}\right)^{3N/2} 
\int {\rm d}x^{3N} \exp(-\beta V(x))
\end{equation}
can be factored into a product of the kinetic part and a product 
depending on the zeros $\cbeta_k = \beta_k +i \tau_k$, with $\cbeta_{-k}=\cbeta_k^*$
of this integral function in the complex temperature plane, 
\begin{equation}
Z(\beta) = \left( \frac{1}{2 \pi \beta}\right)^{3N/2}
 \prod_{k=-M}^M \left( 1-\frac{\beta}{\cbeta_k} \right) \exp\left(
\frac{\beta}{\cbeta_k}\right). \label{partfunc}
\end{equation}
It has been shown that phase transitions can be classified by a set of
three parameters($\alpha, \gamma, \tau_1$), describing the distribution
of zeros close to the real axis, where $\gamma=\tan\nu$ is the crossing
angle between the real axis and the line of zeros, and $\alpha$ is determined
from the approximated density of zeros $\phi(\tau) \sim
\tau^\alpha$ on this line. For infinite systems it has been exactly shown that
$\alpha=0$, $\gamma=0$ and $\tau_1=0$ corresponds to a first order phase
transition, while $\alpha>0$ corresponds to a higher order phase
transition~\cite{Gross1969a}. 
For finite systems $\tau_1$ is always greater than zero
reflecting the size of the system.  The classification
scheme can be extended to values of $\alpha<0$ also being interpreted
as first order phase transitions.  This scheme sensitively reproduces
the space dimension and particle number dependence of the transition
order  in Bose Einstein condensates\cite{Muelken:2000} and the first order nature of the
nuclear multifragmentation phase transition\cite{Muelken:2000b}.
The differences between both schemes can be revealed within a simple statistical
model for atomic clusters.
The multiple normal-modes model \cite{fhb93} describes structural transitions
within small noble gas clusters by considering several isomers and the
vibrational eigenfrequencies of the isomers. For a two isomer system the
partition function can be written as
\begin{eqnarray} Z(\beta)&=&\sum_{i=1}^2
\sigma_i\exp(-\beta E_i) \prod_{j=1}^{3N-6}\frac{2\pi}{\beta\omega_{ij}}
\label{mnm}\\ &=&\beta^{-(3N-6)} \left(\rho_1\exp(-\beta E_1) +
\rho_2\exp(-\beta E_2)\right),\nonumber
\end{eqnarray}
where the $\omega_{ij}$ are the normal modes of isomer $i$ and the
$\sigma_i$ are the permutational degeneracies of the isomers and $\rho_i =
\sigma_i  \prod_{j=1}^{3N-6}\frac{2 \pi}{\omega_{ij}}$.  The zeros of $Z$ 
\begin{equation} \cbeta_k = \left( \ln(\rho_2/\rho_1) + i
(2k+1)\pi \right)/\Delta E, \label{mnm.zeros} 
\end{equation}
lie on a straight
line and are equally spaced yielding $\alpha=\gamma=0$ thus implying
a first order phase transition in any case ($\Delta E = E_2 -
E_1$) (see Figure 1). 
\begin{figure*}
\label{figure1}
\centerline{\includegraphics[clip=,width=18cm]{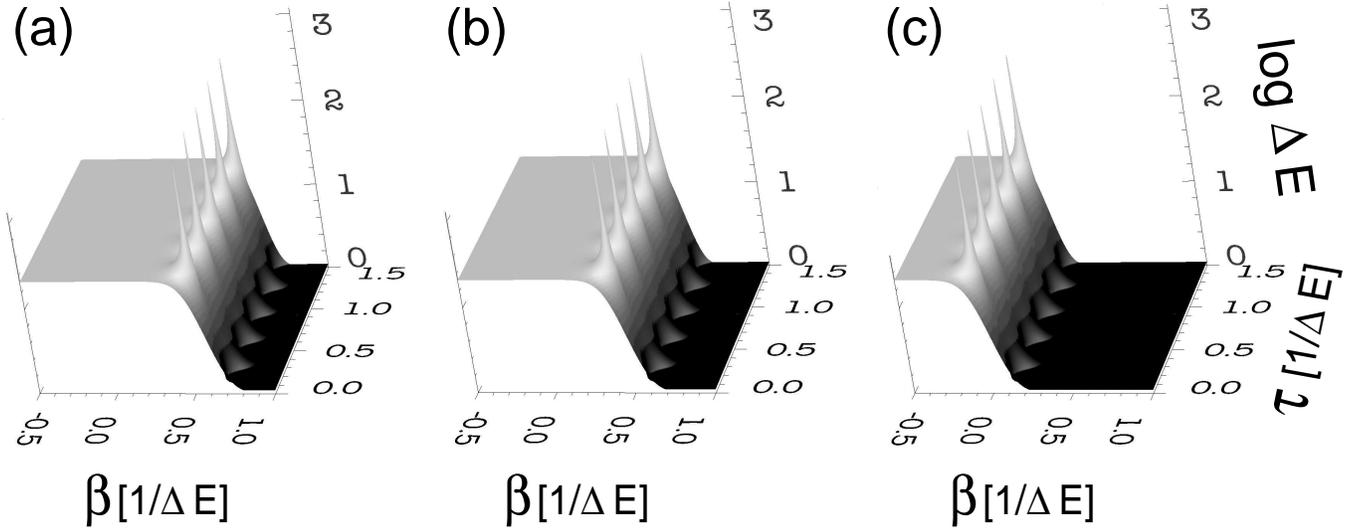}}
\caption{Logarithm of the canonical potential energy difference
expectation value $\log\left(\left<\Delta E\right>\right)$ in the complex
temperature plane for (a) $\rho_2/\rho_1$ = 50000, (b) 
$\rho_2/\rho_1$ = 5000, and (c) $\rho_2/\rho_1$ =0.5. The location
of the zeros of the partition function are signaled by the sharp
needles. In all cases the distributions of zeros indicate first
order phase transitions.}
\end{figure*}
It is important to note, that with increasing
system size the energy difference between the isomers will also
increase, thus $\tau_1$ approaches zero. The microcanonical caloric
curve $T(E)=1/\partial_E\ln(\Omega(E))$ for this model can be
calculated via the inverse Laplace transform $\Omega(E)={\cal
L}^{-1} \left\{ Z(\beta) \right\}$. Fig. 2 shows that the 
back-bending arrogated in the Gross scheme for a first order 
phase transition can be tuned in an out by variation of the 
model parameters.

\begin{figure}
\centerline{\includegraphics[clip=,width=8cm]{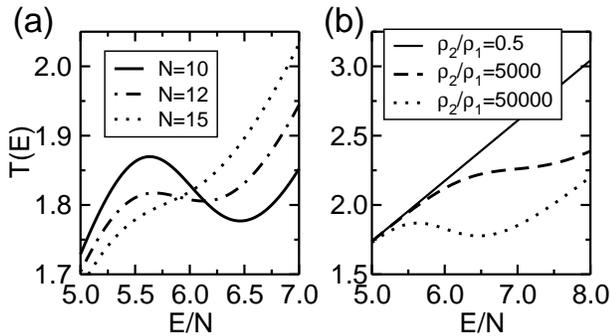}}
\label{figure2}
\caption{Microcanonical caloric curves for the multiple normal-modes model
with energy difference between the isomers $\Delta E=20$.  For {\bf a)}
$N$=10,12, and 15 and constant $\rho_2/\rho_1=50000$  the back-bending is
manifest for $N=10$, can be tuned out by increasing the particle number,
and disappears for $N$ as low as $N=$15. In {\bf b)} for constant $N=10$
the back-bending can be tuned out by decreasing the ratio
$\rho_2/\rho_1$.} \end{figure}

The kinetic part of the partition function $\beta^{-(3N-6)}$ plays the
crucial role. If this is taken into account the micro-canonical caloric
curves change dramatically, whereas this part has no effect on the
distribution of zeros (the particle number dependence of the canonical
partition function is not only reflected by the kinetic part itself but
also implicitly by the ground state energies).  
The change in the topology of the configuration space or
equivalently configurations space regions with significantly
chaniging vibrational entropies seems to 
be a necessary condition for phase transitions in small systems.
Similary results have been  pointed out by Pettini {\sl et al.}. Within 
classical statistical mechanics the kinetic part of canonical
partition function is separable and the partition function splits up
into a kinetic and a potential part which can be handled
independently. Within the microcanonical ensemble structural phase 
transition might be {\sl washed} out or hidden by the kinetic energy 
contributions to the entropy.

\begin{figure}[h]
\label{figure4}
\centerline{\includegraphics[clip=,width=8cm]{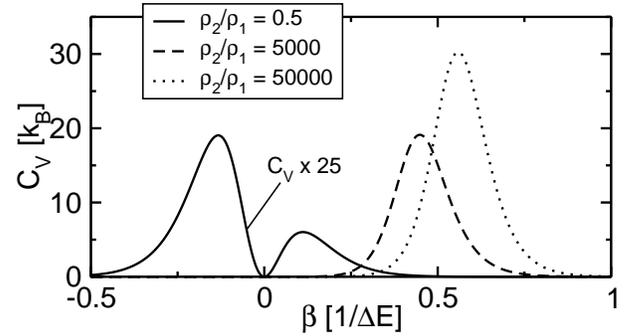}}
\caption{Canonical specific heat reduced by the kinetic contribution for
the same values of $\rho_2/\rho_1$ as in Figure 1. For values with
$\rho_2/\rho_1$ larger than 1 the expected signals of a first order phase
transition are seen. The value $\rho_2/\rho_1$=0.5 corresponds formally
to a first order phase transition at negative temperature. This {\sl
graveling} transition exhibits at positive temperature a very weak hump in
the specific heat (the graph is amplified by a factor of 25).} 
\end{figure}

A very interesting feature of the multiple normal-modes model occurs in the case where
the isomer with the lower ground state energy has a larger vibrational 
entropy (see Fig. 1 (c)). In this case formally 
a first order phase transition at negative temperatures occurs. The 
structural transition between the isomers, which occurs when the
temperature is increased is accompanied by a drop in the 
vibrational entropy. 
This is a {\sl graveling} transition with a
significantly smaller signal in the specific heat than that of the
{\sl normal} transition (see Fig. 3). Fig.3 and Fig.4 show i) that the zeros
in the complex temperature plane sensitively detect phase transitions and ii)
it is very important to use $\beta$ as the natural variable since only this
yields continuous pictures of thermodynamic properties.

\begin{figure}[h]
\label{figure3}
\centerline{\includegraphics[clip=,width=8cm]{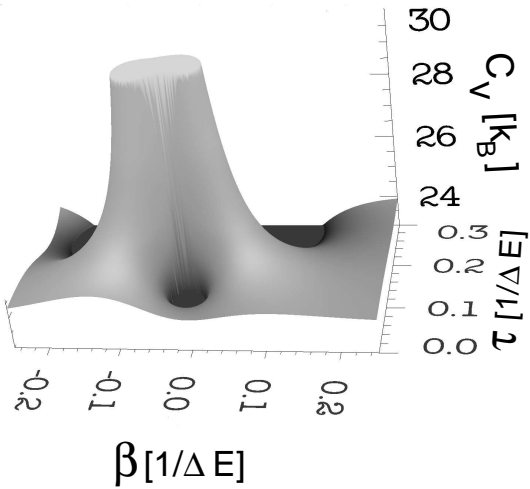}}
\caption{Specific heat in the complex temperature plane
for $\rho_2/\rho_1$=0.5. The figure displays how the interplay of the zero and
the pole of the specific heat influences the behaviour of the specific heat 
curve at positive temperatures.}
\end{figure}

In conclusion we have found that the classification of phase transitions in
small systems based on the curvature of the microcanonical caloric curves
seems to be not rigorous enough to make distinct statements about the order.
We have shown that the investigation of the whole complex temperature plane
adds a valuable amount of information in order to classify phase transitions
in small systems.

\end{document}